\documentclass[usenatbib,useAMS]{mn2e}
\usepackage{times}
\usepackage{epsfig}

\title[LMC self lensing for OGLE-II microlensing observations]{LMC self lensing for OGLE-II microlensing observations}

\author[S. Calchi Novati et al.]
{S.~Calchi Novati$^{1,2,3}$ 
\thanks{E-mail:novati@sa.infn.it}, 
L.~Mancini$^{1,2,3}$, G.~Scarpetta$^{1,2,3}$ and
{\L}.~Wyrzykowski$^{4}$\\
$^{1}$ Dipartimento di Fisica ``E. R. Caianiello'', Universit\`a di Salerno, 
Via S. Allende, 84081 Baronissi (SA), Italy\\
$^{2}$ Istituto Nazionale di Fisica Nucleare (INFN), Sez. di Napoli, Italy\\
$^{3}$ Istituto Internazionale per gli Alti Studi Scientifici (IIASS), Vietri Sul Mare (SA), Italy\\
$^{4}$ Institute of Astronomy, University of Cambridge, Madingley Road, Cambridge CB3 0HA, UK\\
}
\begin{document}

\date{date}

\pagerange{\pageref{firstpage}--\pageref{lastpage}} \pubyear{2009}

\maketitle

\label{firstpage}

\begin{abstract}
In the framework of microlensing searches towards the Large Magellanic Cloud (LMC),
we discuss the results presented by the OGLE collaboration for their
OGLE-II campaign \citep{lukas09}.
We evaluate the optical depth, the duration and the expected rate of events
for the different possible lens populations: both luminous, dominated by 
the LMC self lensing, and ``dark'', the would be compact halo
objects (MACHOs) belonging to either the Galactic or
to the LMC halo. The OGLE-II observational results,  2 microlensing candidate
events located in the LMC bar region with duration of 24.2 and 57.2 days, 
compare well with the expected signal from the luminous
lens populations: $n_\mathrm{exp}=1.5$, with typical duration, for LMC self lensing,
of about 50 days. 
Because of the small statistics at disposal, however,
the conclusions that can be drawn as
for the halo mass fraction, $f$, in form of compact halo objects are not too severe.
By means of a likelihood analysis 
we find an \emph{upper} limit for $f$, at 95\% confidence level, of about 15\%
in the mass range $(10^{-2}-10^{-1})~\mathrm{M}_\odot$ and 26\% 
for $0.5~\mathrm{M}_\odot$.
\end{abstract}


\section{Introduction}
Following the original suggestion of \cite{pacz86,pacz91},
only a few years later, the MACHO, EROS and OGLE collaborations
started their observational microlensing campaign towards the LMC
and the Galactic bulge with the aim of probing, in particular as for the first line of sight,
the Galactic dark matter distribution in form of MACHOs. 
After the first exciting detection of microlensing events,
\citep{macho93,eros93,ogle93}, by now the statistics of detected microlensing candidates
allows one to begin to draw conclusions on the relevant 
scientific issues of interest.
As for the lines of sight towards the LMC,
the MACHO and the EROS collaborations have
presented the final results of their several-years campaigns,
getting to altogether different conclusions as for 
the content of the Galactic halo in compact halo objects.
The MACHO collaboration \citep{macho00} have presented 13-17 microlensing candidate events
and have evaluated an optical depth  $\tau=1.2^{+0.4}_{-0.3}\times 10^{-7}$, far larger
than what is expected from the known possible luminous lens populations.
Their analysis indicates that a mass fraction of about 20\% of the dark matter
halo shoud be composed by MACHOs in the preferred mass range
$(0.15-0.9)~\mathrm{M}_\odot$. 
On the other hand, the analysis of the
EROS collaboration leads to a different conclusion. No viable microlensing
candidate event have been detected, and accordingly
EROS have evaluated a rather tight \emph{upper} limit 
for the fraction of compact halo objects.
In particular they evaluate an upper limit for the optical
depth, due to such lenses, $\tau<0.36\times 10^{-7}$ (95\% confidence level).
This translates, in particular, in an upper limit of about 8\%
for the  halo mass fraction in form of $0.4~\mathrm{M}_\odot$ 
compact halo objects \citep{eros07}.

The strategies followed by the MACHO and the EROS collaborations are different. EROS covered
a larger sky region around the LMC and considered as potential sources
only a subset of clearly identified ``bright'' stars following the 
successful approach
used for Galactic bulge searches \citep{popowski05,hamadache06,sumi06};
whereas MACHO concentrated the observations along 
the LMC bar region, where the contamination to ``MACHO lensing'' 
(lensing events due to lenses belonging to the dark matter halo 
population in form of compact objects) by self lensing
(lensing events with the lens belonging to 
the same luminous population as the source star)
is expected to be larger, and kept as viable sources also blended objects. 
This does not seem to be sufficient, however,
to account for the difference in their results, and indeed
several explanations have been proposed to understand the nature
and the location of the lens for the observed MACHO microlensing candidates.
First, it should be noted that 
it is likely that the lenses do not all belong to the same population \citep{jetzer02}.
In particular, the LMC self lensing has been proposed \citep{sahu94,wu94,gyuk00} 
as a viable solution,
but \cite{mancini04} have shown that 
it can not explain most of the events. \cite{novati06} have studied
the possibility for some of the lenses to belong to the LMC dark matter halo
(a similar possibility, but for observations towards the SMC, have also
been considered by \citealt{dong07}). Finally, two candidates have been acknowledged to belong
to the Galactic disc \citep{alcock01b,kalliva06}.  In fact, the possibility that some of the
microlensing candidates are not microlensing variation at all must be taken
into account. Recently \cite{bennett05} has refined and put up to date the selection
of the MACHO events, taking into account the possible
contamination of variable stars. As a result, 10-12 out of the original set of the 
13 ``set A'' variations \citep{macho00}
are acknowledged as ``likely to be microlensing events'', with a resulting new estimate
of the microlensing optical depth $\tau=(1.0\pm 0.3)\times 10^{-7}$,
still in agreement with the previous one, and such that the overall 
conclusions of this new analysis do not
differ substantially from that of the original MACHO analysis.

In this paper we discuss the recent observational
results presented by \cite{lukas09} for the OGLE-II
campaign towards the LMC. In particular, we carry out
a detailed evaluation of the expected lensing signal
due to the different luminous populations as opposed
to MACHO lensing. The paper is organised as follows.
In Section~\ref{sec:models} we describe the models
for the different lens populations that may give rise
to microlensing events along this line of sight.
In Section~\ref{sec:ana} we present our analysis
based on the evaluation of the microlensing quantities,
the optical depth and the microlensing rate,
and discuss our results as compared to the
observational results of the OGLE campaign.

\section{Models} \label{sec:models}

\subsection{The LMC} \label{sec:lmc}

The LMC luminous components, the \emph{disc} and the \emph{bar}, are modelled
following the analysis presented in a series
of papers by Van der Marel and co-authors \citep{vdm01a,vdm01b,vdm02}. 
The disc and the bar
are considered to be both centered at
$\alpha = 5^{\mathrm{h}} \,
27.6^{\mathrm{m}}$, $\delta =
-\, 69.87^{\circ}$ (J2000) at a distance
of $D=50.1~\mathrm{kpc}$. We assume a bar mass of
$M_{\mathrm{bar}}=1/5\,M_{\mathrm{disk}}$ \citep{sahu94,gyuk00},
with a total visible mass in disk and bar of
$M_{\mathrm{bar}}+M_{\mathrm{disk}}=(2.7\pm 0.6) \times 10^{9}\,
\mathrm{M}_{\sun}$ \citep{vdm02}. The intrinsic shape of the LMC disc is elliptical,
with an inclination angle of $34.7^\circ$. The disc vertical distribution
is described by a $\mathrm{sech}^2$ function, with a flaring
height scale of about $0.3~\mathrm{kpc}$. We use a scale length
for the disc exponential planar distribution of $1.54~\mathrm{kpc}$,
and a boxy shaped bar \citep{zhao00} with length and height scale 
of $1.2~\mathrm{kpc}$ and $0.44~\mathrm{kpc}$, respectively.
Further details are given in \cite{mancini04}.

For the velocity distribution, we assume
a gaussian isotropic profile with line of sight velocity dispersion
of $20.2~\mathrm{km/s}$ \citep{vdm02} for disc stars
(acting both as sources and lenses)
and $24.7~\mathrm{km/s}$ \citep{cole05} for bar stars
(sources and lenses).
A detailed study of the deviation from this 
functional form may be found in \cite{mancini09}.

The lens mass function is a crucial
ingredient as it constitutes the link
between the number of available lenses 
and the overall mass of the luminous component under consideration.
We use \citep{kroupa02,kroupa08} a broken power law
$\xi(\mu_l)\propto \mu_l^{-\alpha}$ with $\alpha=1.3,2.3$ in the mass ranges
$(0.08,0.5)\,\mathrm{M}_\odot$ and $(0.5,1)\,\mathrm{M}_\odot$
respectively ($\mu_l$ is the lens mass). For $\mu_l>1~\mathrm{M}_\odot$ we use 
the present day mass function (PDMF) value $\alpha=4.5$
and we normalize the overall mass in the mass range up
to $120\,\mathrm{M}_\odot$. 
In fact, most of the mass is given by objects with
mass smaller than $1\,\mathrm{M}_\odot$ 
but the use of the PDMF value 
ensure us the correct normalization in each mass range.
The maximum value for the lens mass  should be fixed, once given 
the threshold magnitude for the sources 
(in the present case $I=20.4$, \citealt{lukas09}), 
according to the demand for the lens to be an unresolved object.
Therefore, taking into account the lens distance,
the mass-luminosity relationship and the (varying)
extinction conditions of the specific line of sight \citep{subramaniam05},
we fix this value at  $2\,\mathrm{M}_\odot$ 
(for a resulting average lens mass of $<\mu_l>\sim 0.32~\mathrm{M}_\odot$).
However, we find our results to be quite insensitive to this choice:
moving this limit up to $5\,\mathrm{M}_\odot$ 
the overall expected number of events rises 
only by about 5\%. 
In principle we might include also a sub-stellar component
extending the lens mass range in the brown dwarfs regime
down to $0.01~\mathrm{M}_\odot$. Recent studies \citep{thies_kroupa07}
suggest a flattening in the mass function slope, $\alpha=0.3$,
so that we would find, for the expected number of events,
an increase of only about 4\%.
It is worth recalling that these parametrizations
are based on analyses carried out within our Galaxy. Robust results
from analyses carried out within the LMC itself are still missing,
in particular for stars with mass below $0.4~\mathrm{M}_\odot$
(in fact the most interesting mass range for our purposes)
even if a few first analyses show roughly consistent,
although somewhat steeper, behaviours
\citep{gouliermis06,gouliermis09}. 

A further ``luminous'', even if somewhat elusive, component that can be considered
as a potential lens population is that of the \emph{stellar halo} (SH) of the LMC.
As a fiducial model we include this contribution
following the analysis of \cite{alves04} 
with the parametrization as given in \cite{novati06}.
In particular we attribute to the stellar halo,
endowed with a \emph{spherical} mass distribution,
a total mass, within $8.9~\mathrm{kpc}$ of 
$0.35 \times 10^{9}\,\mathrm{M}_{\sun}$.
Based on a new analysis of OGLE-III RR Lyr{\ae} stars,
\cite{pejcha09} suggested, instead, a \emph{triaxial ellissoid}
mass ditribution for the stellar halo, elongated
along the line of sight to the observer.
As a test model we follow their parametrization,
using the same functional distribution and total mass as in \cite{alves04},
with a position angle of $112.4^\circ$ and an inclination
angle of $6^\circ$, axes ratio $1:2.0:3.5$ and FWHM length
along the line of sight of $7.56~\mathrm{kpc}$.
Since the microlensing rate is proportional,
through the Einstein radius, to the lens-source distance,
the expected signal from the stellar halo
can become relatively large in spite of the overall
small total mass of this component.
For the velocity distribution we use
a gaussian isotropic profile with
line of sight velocity dispersion of $53~\mathrm{km/s}$
\citep{minniti03,borissova06}.

The total dynamical mass of the LMC, 
$8.7\times 10^9\,\mathrm{M}_{\sun}$,
as compared to the discussed luminous components, requires
that more than half of it be comprised in
a dark matter halo component \citep{vdm02}. 
To study the possible contribution of LMC MACHO objects
to the lensing signal we assume a isothermal spherical density profile
with core radius of $2~\mathrm{kpc}$ \citep{macho00}
and a velocity dispersion of $46~\mathrm{km/s}$ \citep{vdm02}.

\subsection{The Milky Way} \label{sec:mw}

Along the line of sight towards the LMC,
the Milky Way (hereafter MW) provides two further luminous lens populations:
the disc and the stellar halo. For the \emph{disc} 
density distribution we follow the parametrisation
of the model discussed in \cite{hangould03} and modified as discussed
in \cite{novati08}. We use a length scale for the exponential profile
of $2.75~\mathrm{kpc}$ and a height scale ($\mathrm{sech}^2$ model including
a flare) of $0.25~\mathrm{kpc}$. Given the mass function as a power law
including the brown dwarf mass range 
we fix the local density in agreement with local star counts 
at $\Sigma_0 \sim 25~\mathrm{M}_\odot$ \citep{kroupa02,kroupa07}. We use a gaussian
isotropic velocity distribution with line of sight
velocity dispersion of $30~\mathrm{km/s}$.
As for the Galactic \emph{stellar halo}
mass function and overall normalization
we follow the analysis of \cite{chabrier03}.
In particular we consider stars up to a mass
of $0.9~\mathrm{M}_\odot$ with
a local normalization of $9.4\times 10^{-5}~\mathrm{M}_\odot~\mathrm{pc}^{-3}$.
We use a mass distribution  with a  $\rho\propto r^{-3}$ radial profile
with flattening 0.6 and line of sight velocity dispersion of $120~\mathrm{km/s}$ \citep{amina08}.

Besides the luminous lens populations,
the largest contribution to the microlensing rate towards the LMC
is expected to come from the would be MACHOs
belonging to the Galactic dark matter halo.
For this component we use the ``standard'' isothermal spherical density profile
with core radius of $5~\mathrm{kpc}$ \citep{macho00},
local density of $7.9\times 10^6~\mathrm{M}_\odot \mathrm{kpc}^{-3}$
and $155~\mathrm{km/s}$ as the value 
for the line of sight velocity dispersion.
For MACHOs (in both halos) we consider a set of delta function
in the mass range $(10^{-5}-10)~\mathrm{M}_\odot$.

Throughout the paper we assume a value for the distance
to the Galactic centre of $8~\mathrm{kpc}$ \citep{trippe08,groenewegen08,gillessen09}.

\section{Analysis} \label{sec:ana}

\subsection{The OGLE-II results}
\cite{lukas09} presented the final results
of the OGLE-II campaign towards the LMC. The monitored fields
are concentrated along the LMC bar region (covering a somewhat
smaller region than the MACHO fields). The observational campaign
lasted 1428 days through 4 years (1996-2000). The search for
microlensing events has been carried out working both with 
a larger set of possible source of even blended objects and with a subsample
of ``bright'' sources only, to which we refer to,
following \cite{lukas09}, as \emph{All} and \emph{Bright} star samples, respectively. 
As a result, from the first set of sources
two microlensing candidate events have been detected,
whose characterictis are summarised in Table~\ref{tab:ogle2},
and none from the second restricted sample. The optical depth
evaluated for the \emph{All}  sample sums up to 
$\tau_\mathrm{obs}=(4.3\pm 3.3)\times 10^{-8}$.

\begin{table}
\caption{Characteristics of the two OGLE-II events observed
towards the LMC (data taken from \citealt{lukas09}).}
\begin{tabular}[h]{c|c|c|c|c}
 event & RA & Dec & $t_\mathrm{E}$ & $\tau \times 10^{-8}$  \\
       & [J2000.0] & [J2000.0] & days &\\
\hline
OGLE-LMC-1 & 5:16:53.26  & -69:16:30.1 & 57.2 & 2.8\\
OGLE-LMC-2 & 5:30:48.00  & -69:54:33.6 & 24.2 & 1.6
\end{tabular}
\label{tab:ogle2}
\end{table}

The observational strategy of this OGLE-II
campaign towards the LMC has been more similar to that
followed by the MACHO collaboration: both for the choice
of the target fields, concentrated along the LMC bar,
where the self-lensing contamination to the MACHO signal is expected
to be larger, as well as for the sample of possible source objects,
with the (delicate) issue of blending to be dealt with.
(Concernig blending, \cite{lukas09} carried
out a thorough analysis, in particular by making use 
of some HST LMC-fields luminosity functions, in order
to correctly evaluate the number of available sources).
Because of this reason, a comparison with the results
of the MACHO collaboration is more straightforward.
However, as we detail below, it turns out
that the results of OGLE are rather in agreement
with those obtained by the EROS collaboration,
which followed a different strategy (monitoring
a much larger area of the sky and considering
a subsample of bright stars for the analysis only).

\subsection{The microlensing quantities}
To analyse the microlensing signal we start from
the evaluation of the microlensing quantities,
the optical depth and the microlensing rate (e.g. \citealt{roulet97}).
We recall that the optical depth allows one to characterise the overall 
lens mass distribution, without giving, however, informations
on any event characteristic (in particular it does not depend
on the lens mass). On the other hand, the microlensing rate
allows one to make estimates on the expected number of events
and their characteristics (in particular of the duration, that is linked
to the lens mass).

\subsubsection{The optical depth}

\begin{table}
\caption{Values of the optical depth ($\tau\times 10^{-8}$) 
towards the centre of the 21 OGLE-II fields, and the two
microlensing events, for the LMC lens populations considered
in the paper: self lensing (SL), stellar halo and MACHO lensing
(dark halo). For stellar halo we report both the results
obtained using the spherical and (in brackets) the ellissoidal model.
RA and DEC are in degree (J2000.0).
}
\begin{tabular}[h]{c|c|c|c|c|c}
 field & RA & DEC& SL & stellar halo & dark halo\\
 1& 83.45& -70.10& 4.36 & 0.591 (1.411)   & 6.01 \\
 2& 82.82& -69.86& 4.45 & 0.621 (1.405)   & 6.09 \\
 3& 82.20& -69.80& 4.65 & 0.651 (1.379)   & 6.20 \\
 4& 81.57& -69.80& 4.78 & 0.678 (1.340)   & 6.32 \\
 5& 80.95& -69.68& 4.63 & 0.679 (1.254)   & 6.37 \\
 6& 80.33& -69.62& 4.37 & 0.671 (1.162)   & 6.40 \\
 7& 79.70& -69.40& 3.89 & 0.629 (1.026)   & 6.33 \\
 8& 79.07& -69.32& 3.44 & 0.599 (0.925)   & 6.29 \\
 9& 78.45& -69.23& 2.98 & 0.564 (0.829)   & 6.24 \\
10& 77.82& -69.15& 2.53 & 0.530 (0.742)   & 6.17 \\
11& 77.17& -69.17& 2.14 & 0.510 (0.681)   & 6.16 \\
12& 76.57& -69.64& 1.50 & 0.611 (0.749)   & 6.91 \\
13& 76.56& -68.72& 1.67 & 0.427 (0.551)   & 5.77 \\
14& 75.95& -69.08& 1.52 & 0.461 (0.565)   & 6.07 \\
15& 75.32& -69.08& 1.28 & 0.445 (0.523)   & 6.07 \\
16& 84.07& -70.16& 4.00 & 0.549 (1.373)   & 5.87 \\
17& 84.70& -70.28& 3.58 & 0.507 (1.314)   & 5.72 \\
18& 85.32& -70.41& 3.12 & 0.466 (1.240)   & 5.58 \\
19& 85.95& -70.58& 2.64 & 0.430 (1.156)   & 5.45 \\
20& 86.57& -70.75& 2.20 & 0.400 (1.072)   & 5.33 \\
21& 80.31& -70.56& 1.95 & 0.954 (1.539)   & 7.88 \\
\hline	     	     	       	            
1& 79.22& -69.27& 3.45 & 0.596 (0.932)  & 6.27 \\
2& 82.70& -69.91& 4.60 & 0.632 (1.410)  & 6.13 \\
\end{tabular}
\label{tab:tau}
\end{table}

\begin{figure}
\includegraphics[width=84mm]{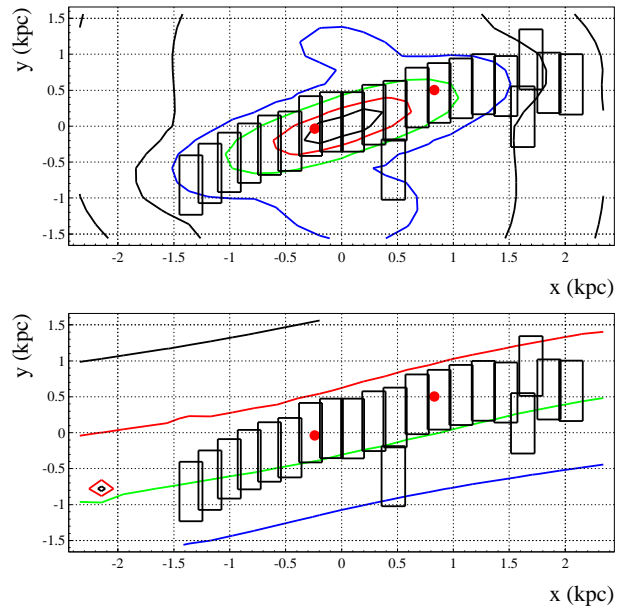}
\caption{Maps of microlensing optical depth towards the LMC: LMC self lensing (upper panel)
and Galactic dark matter halo. The contours indicate the lines of equal optical depth
corresponding to the values $\tau=(1.0,\,1.5,\,2.0,\,3.0,\,4.0,\,4.5)\times 10^{-8}$
(upper panel) and $\tau=(43,\,44,\,45,\,46)\times 10^{-8}$.
The boxes indicate the position of the 21 fields
monitored during the OGLE-II campaigns. The filled circles mark the position
of the two observed OGLE-II candidate microlensing events.
The $x-y$ reference system has its origin at the centre of the LMC,
the $x$-axis antiparallel to the right ascension
and the $y$-axis parallel to the declination.
}
\label{fig:tau}
\end{figure}

Maps for the microlensing optical depth towards the LMC
have been obtained and discussed in a number of papers
(e.g. \citealt{mancini04}). Following the LMC structure,
the LMC self-lensing (LMC disc or bar sources
\emph{and} lenses) optical depth is symmetric around
the LMC centre and  peaks at $\tau = 5\times 10^{-8}$.
Most of the area covered by the OGLE-II fields is within
the line of equal optical depth above $\tau = 2\times 10^{-8}$.
The Galactic dark matter halo optical depth, on the other hand,
is a very smoothly varying function of the position
across the LMC, with the lines of equal optical depth
roughly parallel to the LMC bar axis, and typical value within the OGLE-II fields 
of $\tau = (44-45)\times 10^{-8}$. In Fig.~\ref{fig:tau} we show the
maps of the optical depth for these two lens populations together
with the location of the OGLE-II fields and the position of the observed events.
As for the Galactic disc optical depth, this also turns out
to be a smoothly varying function of the position,
with typical value along the LMC bar down to 10 times
as small as the LMC self lensing, with typical value
of $\tau \sim 0.40 \times 10^{-8}$. On the other hand, the LMC dark matter halo
optical depth \citep{mancini04} shows a strong variation with the position,
also due to the inclination of the LMC disc.
Along the LMC bar it results $\tau \sim 6\times 10^{-8}$.
In Table~\ref{tab:tau} we report the values
of the optical depth for all the LMC lens populations 
considered along the lines of sight towards
the 21 OGLE-II fields and the two microlensing candidate events.
The LMC stellar halo optical depth is strongly
enhanced, in particular towards the central LMC region,
when using the \cite{pejcha09} triaxial model as opposed
to the spherical one of \cite{alves04}.
Nevertheless, this contribution remain small
with respect to the LMC self lensing.
As for the Galactic stellar lensing, the contribution
of the disc and the stellar halo turn out to be rather negligible
with respect to the LMC self lensing, and this holds in
particular because all of the observed fields
are concentrated along the LMC bar region.
For the Galactic disc and stellar halo we find 
$\tau \sim 0.4\times 10^{-8}$ and
$\tau \sim 0.2\times 10^{-8}$, respectively.

As for the line of sight towards the OGLE-II observed events, 
both of them happen to fall within 
the line of equal LMC self-lensing optical depth
of $\tau= 3.0\times 10^{-8}$ (upper panel of Fig.~\ref{fig:tau}),
with the innermost event within that of $\tau= 4.5\times 10^{-8}$. These values compare well
with the overall observed estimate $\tau_\mathrm{obs}=(4.3\pm 3.3)\times 10^{-8}$.

\subsubsection{The microlensing rate}

We have evaluated the microlensing rate along the lines of sight
towards the 21 fields observed during the OGLE-II campaign
for the different lens populations we consider.
We have made use of the detection efficiency evaluated as described
in \cite{lukas09} for both the \emph{All} and the \emph{Bright} samples of sources
to take into account of the observational effects.
Coherently with the analysis of \cite{lukas09}, we reduce the 
efficiency by a factor 0.9 to take into account binary systems
that might be missed by the pipeline. 
To evaluate the differential microlensing rate
we have included, beside the random component of the velocity, 
the bulk motion of the LMC and of the observer and the drift
velocity of the LMC disc following the scheme outlined in \cite{novati06}.

\begin{table}
\caption{Results of the microlensing rate analysis for the 
different lens populations we consider: evaluation
of the duration and number of expected events, $n_\mathrm{exp}$ as evaluated
on the 21 fields monitored during the OGLE-II campaign.
For the Einstein time we report the values averaged across
the observed fields. For $n_\mathrm{exp}$ we report the sum
over the observed fields, taking into account the efficiency,
for both the \emph{All} and the \emph{Bright} sample of sources.
SL and SH stand for self lensing and stellar halo, respectively.}
\begin{tabular}[h]{c|ccc|ccc}
lenses&\multicolumn{3}{c}{$t_\mathrm{E}$}&\multicolumn{2}{c}{$n_\mathrm{exp}$}\\
 & 16\% & 50\% & 84\% & \emph{All} & \emph{Bright} \\
\hline
LMC SL & 26. & 52. & 100. & 1.1 & 0.46\\
LMC SH & 24. & 46. & 83. & 0.20 & 0.086\\
MW disc & 18. & 34. & 65. & 0.12 & 0.056\\
MW  SH & 19. & 34. & 65. & 0.071 & 0.031\\
\hline
MW halo &&&&&\\
$10^{-5}~\mathrm{M}_\odot$ & 3.2 & 4.7 & 11. & 0.40 & 0.19\\
$10^{-4}~\mathrm{M}_\odot$ & 3.3 & 4.8 & 11. & 3.3 & 1.6\\
$10^{-3}~\mathrm{M}_\odot$ & 3.5 & 5.5 & 12. & 12. & 5.0\\
$10^{-2}~\mathrm{M}_\odot$ & 6.3. & 9.7 & 18. & 21. & 9.2\\
$0.1~\mathrm{M}_\odot$ & 13. & 20. & 36. & 27. & 12.\\
$0.5~\mathrm{M}_\odot$ & 25. & 40. & 69. & 18. & 7.7\\
$1.0~\mathrm{M}_\odot$ & 33. & 56. & 94. & 14. & 6.0\\
$10~\mathrm{M}_\odot$ & 91. & 150. & 230. & 4.3 & 2.0\\
\hline
LMC halo &&&&&\\
$10^{-5}~\mathrm{M}_\odot$ & 3.2 & 4.7. & 11. & 0.15 & 0.071\\
$10^{-4}~\mathrm{M}_\odot$ & 3.3 & 4.9 & 11. & 1.1 & 0.48\\
$10^{-3}~\mathrm{M}_\odot$ & 3.6 & 6.5 & 11. & 2.1 & 0.91\\
$10^{-2}~\mathrm{M}_\odot$ & 9.3 & 14. & 23. & 3.9 & 1.7\\
$0.1~\mathrm{M}_\odot$ & 22. & 35. & 57. & 2.9 &1.2\\
$0.5~\mathrm{M}_\odot$ & 47. & 74. & 120. & 1.5 &0.68\\
$1.0~\mathrm{M}_\odot$ & 63. & 100. & 160. & 1.1 & 0.50\\
$10~\mathrm{M}_\odot$ & 160. & 250. & 340. & 0.22 & 0.10\\
\end{tabular}
\label{tab:rate}
\end{table}

In Table~\ref{tab:rate} we report the main result of the present analysis
on the expected microlensing signal:
the median value of the duration and the number 
of  events, $n_\mathrm{exp}$. As for the \emph{All} sample of sources, 
the expectations for $n_\mathrm{exp}$  from the luminous
lens populations sum up roughly to 1.5 events, with more than 70\%
of the lensing signal expected from the LMC self lensing. 
This compares well with the observed number
of events $n_\mathrm{obs}=2$. The prediction for the
number of expected events for the restricted bright sample,
$\sim~0.65$ events, is also fully compatible with the null result
obtained in this case. As it follows from the location of the observed fields,
we can also see that the expected signal from the Galactic disc
is significantly smaller as compared to the LMC self-lensing one.

For the LMC stellar halo lensing we have considered
here the fiducial spherical \cite{alves04} model. In fact the number of expected
events using the \cite{pejcha09} triaxial model increases significantly,
by about 50\%. However, this contribution remains small with respect to 
that of the LMC self lensing, so that
the overall increase is of only about 6\% and our overall conclusion
are not significantly changed when using one model or the other.

Analogously, one might consider the same triaxial spatial distribution,
elongated along the line of sight, 
also for the dark matter LMC halo. As we have shown
to be the case for the stellar one, we may expect the
optical depth, and correspondingly the microlensing rate,
to increse significantly, mainly because of the increased
mean lens-source distance. However, since 
the ratio of LMC over
Galactic compact halo objects rate is quite small, less
than 10\% (Table~\ref{tab:rate}), we do not expect our results to be significantly
affected by this choice. This is the case also 
because of the small statistics of events at disposal.
A larger set would allow one a study of the spatial
distribution, too, and in that case a different
spatial distribution for the LMC halo contribution
might become a relevant issue \citep{novati06}.

The expected duration varies 
across the observed fields for the LMC self lensing, with larger
values expected moving from the LMC centre, with variations up to about 10\%
\citep{mancini04}. In Table~\ref{tab:rate} we have reported the mean 
values of the duration across the 21 OGLE fields.
In particular, as for the lines of sight towards the two candidates
it results a median (average) value  $t_\mathrm{E}=36\,(49)~\mathrm{days}$
and $t_\mathrm{E}=47\,(61)~\mathrm{days}$,
to be compared with the observed values $t_\mathrm{E}=57.2~\mathrm{days}$
and $t_\mathrm{E}=24.2~\mathrm{days}$, for OGLE-LMC-I and OGLE-LMC-II,
respectively.

\begin{figure}
\includegraphics[width=84mm]{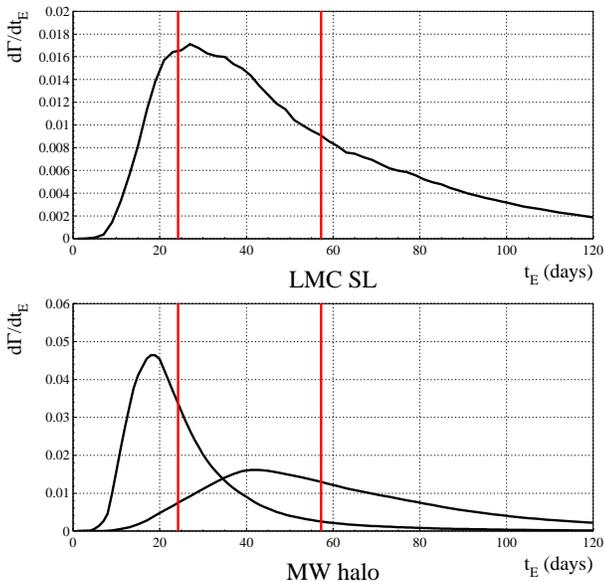}
\caption{Normalised microlensing differential rate 
$\mathrm{d}\Gamma_\mathcal{E}/\mathrm{d}t_\mathrm{E}$
towards the LMC corrected for the OGLE-II efficiency: 
LMC self lensing (upper panel)
and Galactic dark matter halo (0.1 and 1 M$_\odot$, the former peaked
at lower values of the duration). The vertical solid lines
indicate the value of the duration of the two observed OGLE-II events.
}
\label{fig:rate}
\end{figure}

In Fig.~\ref{fig:rate} we show the normalised differential rate
$\mathrm{d}\Gamma_\mathcal{E}/\mathrm{d}t_\mathrm{E}$, corrected for the efficiency
$\mathcal{E}(t_\mathrm{E})$,
for LMC self lensing and for Galactic dark matter halo lenses of $0.1~\mathrm{M}_\odot$
and $1~\mathrm{M}_\odot$.
Superimposed on the differential rate we show the values
of the evaluated duration
for the two events observed using the \emph{All} sample of sources.
Comparing the expected durations with the observed ones
we can see, as for self lensing, that the evaluated duration of OGLE-LMC-1
agrees very well with the median value, and, though somewhat short,
that of OGLE-LMC-2, is still fully compatible.
As for the MW MACHO populations, the observed durations
are compatible with those expected for MACHO masses in the mass range
$(0.1-1)~\mathrm{M}_\odot$.

The expected number of events for the dark populations
(both MW and LMC halos) is shown, as a function
of the MACHO mass, in the upper panel of Fig.~\ref{fig:likeli}.
In particular, for a full halo of $0.05~\mathrm{M}_\odot (0.5~\mathrm{M}_\odot)$ compact objects
we expect, for the MW halo, $n_\mathrm{exp}=28$ and $n_\mathrm{exp}=12$ 
($n_\mathrm{exp}=18$ and $7.7$)
for the \emph{All} and the \emph{Bright} sample of sources, respectively,
whereas for the LMC halo we evaluate $n_\mathrm{exp}=3.5$ and $n_\mathrm{exp}=1.5$
($n_\mathrm{exp}=1.5$ and $0.68$).
For both MW and LMC, the microlensing rate
is severely suppressed by the sharp decline
of the efficiency function $\mathcal{E}(t_\mathrm{E})$ \citep{lukas09}
for small values of $t_\mathrm{E}$: this explains 
the plateau reached by the expected duration values,
as well as the sharp decline for $n_\mathrm{exp}$, 
for compact halo object masses below 
$10^{-3}~\mathrm{M}_\odot$.

\subsection{The mass halo fraction in form of MACHOs}

The evaluation of the microlensing rate together
with the result of the OGLE-II observational campaign
as for the number of observed microlensing events
allows us to put some constraints on the mass halo fraction
in form of MACHOs, $f$. In particular we perform a likelihood
analysis with
\begin{equation} \label{eq:like}
L\left(f,m\right) = \exp\left(-N_\mathrm{exp}(f,m)\right) 
\prod_{i=1}^{N_\mathrm{obs}}\left.\frac{\mathrm{d}
{\Gamma}_{i,\mathcal{E}}}{\mathrm{d}t_\mathrm{E}} 
\right|_{t_{\mathrm{E},event}}\!\!\!\!\!\!\!\!\!\!\!\!\!\!\!\!,
\end{equation}
where, for the differential rate, we sum up
over all the luminous and dark populations we consider.
The halo mass fraction, $f$, enters as a 
multiplicative factor of the differential rate
and of the expected number of events in front
of the contribution of the dark populations
(MW and LMC halos).

\begin{figure}
\includegraphics[width=84mm]{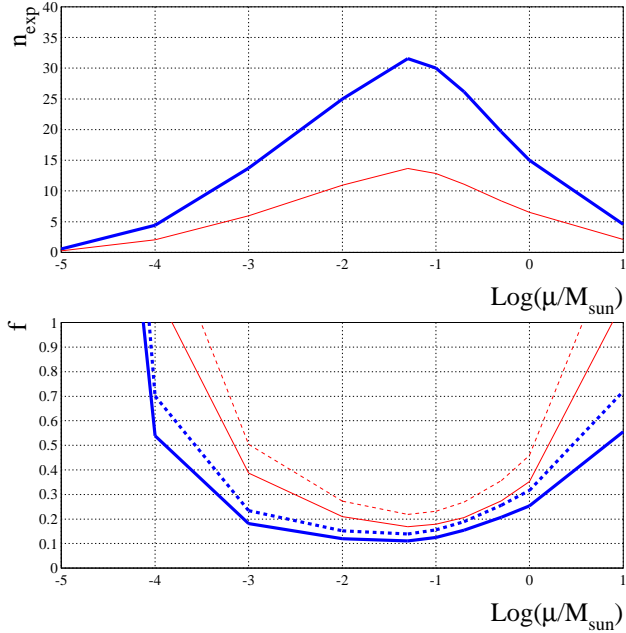}
\caption{Upper panel: number of expected events for (full) dark matter halos
(MW and LMC). Bottom panel: as a result of the likelihood analysis,
90\% and 95\% confidence level (solid and dashed lines, respectively) upper limits
for the halo mass fraction in form of compact halo objects.
The tick and thin lines indicate the results
for the \emph{All} ($n_\mathrm{obs}=2$) and the \emph{Bright} ($n_\mathrm{obs}=0$) 
sample of sources, respectively.
The number of expected events from the luminous populations
is $n_\mathrm{exp}=1.5$ and $n_\mathrm{exp}=0.65$, respectively.
}
\label{fig:likeli}
\end{figure}

Keeping the MACHO mass value fixed as a parameter
we can evaluate upper and lower limits for the 
halo mass fraction $f$. In Fig.~\ref{fig:likeli} (bottom panel)
we display the 90\% and 95\% confidence levels \emph{upper} limits
for $f$, for both the \emph{All} sample and the \emph{Bright} sample
(thick and thin lines, respectively).
As noted in the previous Section,
the number of  expected events from the luminous lens
components is compatible with the observed one
in both cases, therefore we do not show the corresponding (extremely small)
lower limit. Corresponding to the maximum in the number of
MACHO events (upper panel, Fig.~\ref{fig:likeli}),
we find the tighter constraints for the halo
component in form of MACHOs in the mass range
$(10^{-2}-10^{-1})~\mathrm{M}_\odot$.
For $0.05~\mathrm{M}_\odot$ MACHOs we find,
at the 90\% (95\%) confidence levels,
$f<11\% (14\%)$  for the \emph{All} sample
and $f<17\% (22\%)$ for the \emph{Bright} sample.
Finally, for a MACHO mass of $0.5~\mathrm{M}_\odot$, about the value preferred
by the MACHO collaboration analyses, we find
$f<21\% (26\%)$ (\emph{All} sample) and $f<27\% (36\%)$ (\emph{Bright} sample).

We may compare these results  with those obtained under the assumption of 
$n_\mathrm{obs}=0$ with a confidence
limit analysis based on a Poisson distribution
for the number of expected MACHO lensing events 
(as that carried out in \citealt{lukas09}).
The analysis  yields the same results
for the \emph{Bright} sample ($n_\mathrm{obs}=0$)
but a few differences arise for the \emph{All} sample
($n_\mathrm{obs}=2$), where we are including the information
on the observed events, and in particular
of their timescale, and that on the expected number
of events from the luminous populations.
For the \emph{All} sample we expect the limits to rise
whenever the observed durations fit well the expected ones
for the dark matter components.
For MW compact halo objects 
this holds in the mass range $(10^{-1}-1)~\mathrm{M}_\odot$ (Table~\ref{tab:rate}, 
Fig.~\ref{fig:rate}).
In particular, for a dark matter
halo of compact objects of $0.5~\mathrm{M}_\odot$ with the simpler
Poisson analysis we obtain $f<12\% (15\%)$ at the 90\% and 95\% 
confidence levels\footnote{The results we obtain are somewhat
different from those reported in \cite{lukas09}, where
the analysis was based on a few approximations.
In particular for $0.4~\mathrm{M}_\odot$ MACHOs we evaluate an upper limit
$f<14\%$ at 95\% confidence level.}, 
respectively, to be compared with $f<21\% (26\%)$.

\begin{figure}
\includegraphics[width=84mm]{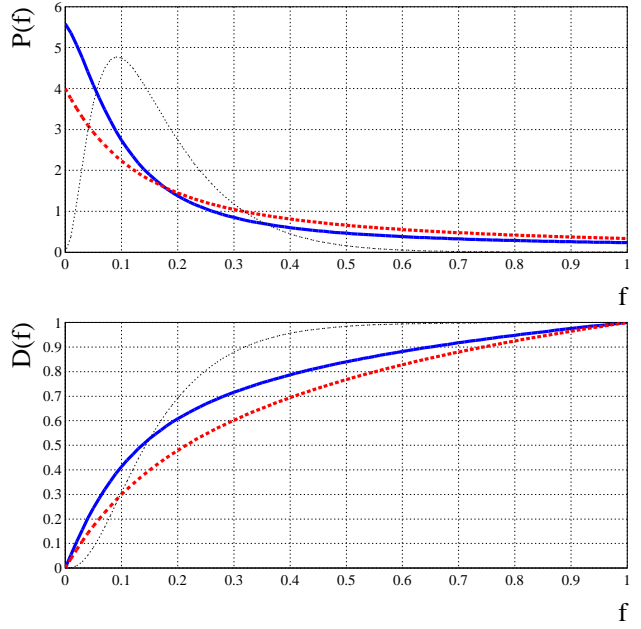}
\caption{Probability (upper panel) and cumulative distribution
for $f$, the halo mass fraction in form of MACHOs,
obtained after integration on the MACHO mass
with a logaritmic flat prior of the likelihood function.
The solid (dashed) lines are for the \emph{All} ($n_\mathrm{obs}=2$) 
and the \emph{Bright} ($n_\mathrm{obs}=0$) sample of sources,
respectively. The (thinner) dotted line, for the \emph{All} sample,
is obtained neglecting the contribution to the microlensing rate
of the luminous populations, namely, assuming that the
observed events \emph{are} due to MACHOs.
}
\label{fig:probf}
\end{figure}

Finally, we can integrate over the mass
the expression of the likelihood 
in the full range considered ($(10^{-5}-10)~\mathrm{M}_\odot$)
and, by Bayesian inversion, obtain
the one dimensional probability
ditribution for the halo mass fraction, $P\left(f\right)$.
Here we use a $\mathrm{d}\log(\mu_l)$ prior on the mass
(but the results are similar, with a resulting
distribution just somewhat broader, for a flat
prior $\mathrm{d}(\mu_l)$).
We show the halo mass fraction in form
of MACHOs probability distribution, $P(f)$, together
with the corresponding cumulative distribution, $D(f)$,
in Fig.~\ref{fig:probf}.

Although, as discussed, the expected microlensing
signal contributed by the luminous lens populations
is sufficient to explain the observed events,
in principle it is possible to take 
the opposite view, namely that the luminous populations
\emph{do not} contribute to the microlensing rate.
In this case we may evaluate, for the \emph{All} sample,
what halo fraction is implied by the two observed events,
and it therefore makes sense to evalute
both an upper and a lower limit for $f$.
This can be looked at as the extreme case 
(opposite to the Poisson-based analysis with $n_\mathrm{obs}=0$)
that fixes  the  limits that can be put on the MACHO contribution.
The resulting probability distribution for $f$, together
with the cumulative distribution, is shown
in Fig.~\ref{fig:probf}: the halo fraction
peaks at $f\sim 10\%$ with a dispersion
of about $8\%$. In particular, for a $0.5~\mathrm{M}_\odot$
compact halo objects halo we find lower and upper
limits for $f$ of $4\%$ and $32\%$ (95\% confidence level).

\section{Conclusion}

In this paper we have analysed
the recent results presented by the OGLE collaboration
as for their OGLE-II campaign directed towards
the bar region of the LMC \citep{lukas09}.
In particular we have carried out a detailed
analysis of the expected signal from
the luminous lens populations (LMC bar, disc and stellar halo,
MW disc and stellar halo) as compared to MACHO lensing
(LMC and MW dark matter halos).

The overall conclusion of our analyses,
in agreement with that reported in \cite{lukas09},
is that the observed signal is in full agreement 
with that expected from the luminous components
(largely dominated by the LMC self lensing). 
This holds looking both at the duration
and at the number of the observed events. The relatively small statistic
at disposal, however, does not allow at the same time
to put very strong constraints on the contribution
to the halo in form of compact halo objects. Indeed the evaluated upper limit 
for the halo mass fraction in form of MACHOs, $f$,
is still, even if only marginally, consistent
with the positive MACHO signal reported by the MACHO collaboration.
On the other hand, it is
much less severe than that reported by the EROS collaboration.

In particular, the estimated value for the optical depth
for the two observed events
$\tau_\mathrm{obs}=(4.3\pm 3.3)\times 10^{-8}$,
compares well with our estimate of $\tau$ for 
LMC self lensing: both events lie within 
the line of equal optical depth $\tau= 3.0\times 10^{-8}$,
with the innermost one within that of $\tau= 4.5\times 10^{-8}$.
Furthermore, through the evaluation of the microlensing rate, we have shown
that the observed signal, 2 events with duration
in the range 20-60 days, is compatible with 
the expected signal from the luminous lens components,
with $n_\mathrm{exp}=1.5$ and a typical duration,
for LMC self lensing, of 50 days.
As for the halo mass fraction in form of MACHOs,
given the agreement of the observed signal
with self lensing, we report an \emph{upper} limit only.
Through a likelihood analysis we have evaluated
an \emph{upper} limit for $f$ at 95\% 
confidence level as low as $14\%$ for compact objects 
of $0.05~\mathrm{M}_\odot$ and values varying in the range 16\%-32\%
in the mass range $(0.1-1)~\mathrm{M}_\odot$, that
preferred by the analyses of the MACHO group 
(and for which the expected durations of the would be Galactic MACHO populations
are in agreement with the observed ones). 

The OGLE-II campaign sampled the bar region of the LMC only. 
Together with the low statistics of observed events 
this makes more difficult the task of disentagling the would be
MACHO lensing from the lensing signal due to the luminous lens populations.
In fact, a larger number of observed events is crucial to this purpose
as it may allow one analyses such those carried out
in \cite{mancini04} and \cite{novati06} on the duration
and the spatial distribution of the events.
The upcoming results of the OGLE-III campaign, 
together with those of SuperMACHO \citep{rest05}, which both covered
a larger area of the sky with an expected larger statistics
of events, should eventually allow one
to put further constraints on the distribution
of the dark matter halo in form of MACHOs.

\section*{Acknowledgments}
SCN acknowledges support from the Italian Space Agency (ASI).
SCN, LM and GS acknowledge support by FARB-2009 of
the University of Salerno.
{\L}W acknowledges generous support from the European Community's
FR6 Marie Curie Programme, Contract No. MRTN-CT-2004-505183
``ANGLES'' and EC FR7 grant PERG04-GA-2008-234784.

\bibliographystyle{mn2e}

\bibliography{ogle}

\label{lastpage}

\end{document}